\documentclass[useAMS,usenatbib]{mn2e}

\usepackage{amsmath,amsfonts,epsfig,subfigure,natbib}

\title[{\em XMM-Newton\/} Surveys of the CFRS Fields - III]{{\em
XMM-Newton\/} Surveys of the Canada-France Redshift Survey Fields -
III: The Environments of X-ray Selected AGN at $0.4<z<0.6$}
\author[Waskett et al.]
{T. J. Waskett$^1$\thanks{E-mail: Tim.Waskett@astro.cf.ac.uk},
S. A. Eales$^1$,
W. K. Gear$^1$,
H. J. McCracken$^2$,
S. Lilly$^3$, 
\newauthor 
M. Brodwin$^4$ \\ 
$^1$Department of Physics and Astronomy, University of Wales Cardiff,
PO Box 913, Cardiff, CF24 3YB, UK\\
$^2$Institut D'Astrophysique de Paris, 98bis, bd Arago -75014 Paris, France\\
$^3$ETH Hoenggerberg Campus, Physics Department, HPF G4.1, CH-8093,
Zurich, Switzerland\\
$^4$Jet Propulsion Laboratory, California Institute of Technology,
M/S 169-506, 4800 Oak Grove Drive, Pasadena, CA 91109}

\begin{document}

\maketitle

\begin{abstract}
The environmental properties of a sample of 31 hard X-ray selected AGN
are investigated, from scales of 500 kpc down to 30 kpc, and are
compared to a control sample of inactive galaxies.  The AGN all lie in
the redshift range $0.4<z<0.6$.  The accretion luminosity-density of
the Universe peaks close to this redshift range, and the AGN in the
sample have X-ray luminosities close to the knee in the hard X-ray
luminosity function, making them representative of the population
which dominated this important phase of energy conversion.

Using both the spatial clustering amplitude and near neighbour counts
it is found that the AGN have environments that are indistinguishable
from normal, inactive galaxies over the same redshift range and with
similar optical properties.  Typically, the environments are of
sub-cluster richness, in contrast to similar studies of high-z
quasars, which are often found in clusters with comparable richness to
the Abell R$\geq 0$ clusters.

It is suggested that minor mergers with low mass companions is a
likely candidate for the mechanism by which these modest luminosity
AGN are fuelled.
\end{abstract}

\begin{keywords}
galaxies: active - X-rays: galaxies - galaxies: evolution
\end{keywords}

\section{Introduction}
Studying the environments of AGN has many motivations, including
providing constraints on galaxy evolution models and how well AGN
trace the normal galaxy distribution.  The motivation in this work,
however, is to determine the relative importance of various possible
fuelling mechanisms that could power the activity in a central engine.
For example, does the Mpc scale environment of a galaxy induce the AGN
phenomenon somehow, or is the region immediately surrounding the
supermassive black hole (SBH) in a galactic nucleus responsible for
the onset of an AGN burst?  Although discovering the exact details of
the fuelling mechanism of any given AGN is beyond this study, studies
of environment can certainly help to narrow down the possibilities
from the vast array of theoretical models proposed thus far.  By
comparing the environments of AGN against a control sample of
otherwise normal galaxies, differences may indicate a fundamental
property of galactic nuclei that causes them to be active rather than
inactive.

This is the third in a series of papers based on {\em XMM-Newton\/}
data of the Canada-France Redshift Survey fields (CFRS).  The first
\citep[][hereafter paper 1]{2003MNRAS.341.1217W} concentrated on the
X-ray sub-mm relation in the 3-h and 14-h CFRS fields, while the
second \citep[][hereafter paper 2]{2004MNRAS.350..785W} presented the
source catalogues along with the optical properties and redshift
distribution of the X-ray sources in those same fields, which were
largely photometric redshifts derived from the $UBVI$ optical
catalogues of the Canada-France Deep Fields survey
\citep[CFDF,][]{2001A&A...376..756M}.

We assume an $H_{0}$ of 75 km s$^{-1}$~Mpc$^{-1}$ and a concordance Universe 
with $\Omega_{M}=0.3$ and $\Omega_{\Lambda}=0.7$.

\subsection{AGN Fuelling Mechanisms}
As \citet{1998ApJ...495..152L} put it, there are three distinct phases
in the process of fuelling an AGN: phase 1, gas must be moved from the
galactic scale into the central few hundred pc; phase 2, the
instabilities of a self gravitating disc further compact the gas until
it forms an accretion disc around the central SBH; and finally,
accretion processes in the disc itself allow the gas to be either
swallowed by the SBH or ejected along the rotation axis.  One of the
reasons for studying the environments of AGN is to understand the
first phase, by which the fuel supply is made available to the SBH
through gas transport on galactic scales.  For this reason we shall
ignore smaller scale processes, such as disc instabilities and
galactic bars, to mention just two of the many possibilities suggested
for the second fuelling mechanism phase.

Various mechanisms proposed in the literature lead to definite
predictions about the nature of the environments of AGN.  We briefly
summarise some of those mechanisms and predictions here:

\begin{itemize}

\item Interactions/major mergers:  This model involves two comparably
sized galaxies interacting though their mutual gravitational
attraction, leading to the amalgamation of the two central SBHs with
large quantities of gas being driven inwards to fuel the resulting
central engine \citep[e.g.][]{2000MNRAS.311..576K}.  The QSO produced
from this will reside in either a massive elliptical host (due to the
disruption of the original merging galaxies and subsequent relaxation
of the stellar population), or a highly irregular one depending on the
time-scale for the onset of the nuclear activity, most likely in a high
density environment where mergers are more common.  The model also
correctly predicts the observed space density of QSOs that is seen to
peak at $z\sim 2$.  Although a successful model in predicting both the
morphology and environments of very luminous QSOs
\citep[e.g.][]{1999MNRAS.308..377M,2001MNRAS.321..515M} it does not
explain fainter AGN that are found in spirals as well as ellipticals.
If major mergers were responsible for lower luminosity AGN then many
more host galaxies should be observed to have disturbed morphology or
signs of recent interactions, which does not seem to be the case
either \citep[e.g.][]{2003ApJ...595..685G}.

\item Minor mergers:  For AGN of lower than QSO luminosity, such as
Seyferts, it has been proposed that mergers of small companion
galaxies (SMC or smaller) may induce nuclear activity in gas rich hosts
\citep{1998ApJ...496...93D}.  This is particularly relevant to the
study here as we specifically concentrate on lower luminosity AGN (see
section~\ref{AGN_sample}).  Predictions of this model include
undisturbed hosts and no need for significantly different environments
(in terms of bright galaxies) from those of comparable field galaxies.
However, detecting such small companions around anything other than a
nearby galaxy is problematic, so a direct observation of a high
frequency of small companions around high redshift AGN is unlikely to
be made any time soon.
 
\item Harassment:  Originally proposed by \citet{1996ApJ...457..455M}
to explain the morphological evolution of galaxies in rich clusters,
it has also been suggested as an AGN fuelling mechanism
\citep{1998ApJ...495..152L}.  The mechanism consists of
numerous high speed interactions that a relatively small disc galaxy
experiences while travelling through a cluster.  Rather than the
cataclysmic, but relatively slow, interactions experienced by two
galaxies undergoing a merger in the field, or in the centre of a
cluster, the higher speeds at which galaxies fly past one another in
the outskirts of a cluster cause a member galaxy to be jiggled around
but otherwise remain largely unchanged.  Dynamical instabilities
induced by this ``galaxy harassment'' channel gas into the central few
kpc of sub-$L_{*}$ galaxies, where it becomes available as fuel for an
AGN.  Relatively new additions to the cluster, i.e. In-falling
galaxies, are more likely to have large gas reservoirs and so are more
likely to host an AGN.  Clearly, predictions of this mechanism include
the presence of a relatively rich cluster, with the AGN either in the
periphery or in the process of falling in towards the cluster.
\citet{2004MNRAS.347.1241S} present some evidence that this may be the
case for low-$z$ quasars, with nearly half of their $z<0.3$ sample
being found within $1-3~h^{-1}$ Mpc of a cluster centre.  The
harassment model also predicts that hosts should show slightly
disturbed morphologies but not be totally disrupted.

\item Cooling flows:  Clusters of galaxies contain a hot intra-cluster 
gas that tends to be many times as massive as the cluster galaxies
themselves.  This provides a potentially huge reservoir of fuel for an
AGN residing in a central cluster galaxy \citep{1986ApJ...305....9F},
if the gas were to experience a radiative loss of energy and hence
fall in towards the centre of the cluster -- a cooling flow.  Again,
clear predictions can be made for this mechanism, such as a high
relative fraction of AGN found in clusters undergoing cooling flows.
However, this mechanism can only applies to AGN in central
cluster/group galaxies and does not explain the AGN found in galaxies
away from the cluster centre \citep{1997AJ....113.1179H}.

\end{itemize}

The above list is by no means exhaustive but it gives a brief example
of the variety of theoretical models on offer to explain the AGN
phenomenon.  Of course many mechanisms proposed to fuel AGN can be
equally applied to a nuclear starburst and in reality different
mechanisms are likely to be more important for different classes of
AGN.  As always, it is a complex problem without a single simple
answer.

\subsection{Previous Work}
In general, most of the previous investigations into the environments
of AGN concern optically or radio selected QSOs.  Radio-loud QSOs are
now almost universally acknowledged to lie in over-dense regions,
typically clusters of Abell 0/1 richness, across a large range in
redshift \citep[e.g.][]{2000MNRAS.316..267W,2001MNRAS.321..515M,
2003MNRAS.346..229B}; whereas there is still some disagreement over
whether the same is true for radio-quiet QSOs.
\citet{2001MNRAS.323..231W} and \citet{2001MNRAS.321..515M} find no
significant difference between the environments of matched samples of
radio-loud and radio-quiet QSOs while \citet{2000MNRAS.313..252S},
amongst others, claim that radio-quiet QSOs are no more likely to be
found in rich environments than non-active galaxies.  However,
differences between the various techniques and survey designs employed
by different workers are likely to play some part in the
discrepancies.

\citet{2004MNRAS.347.1241S} employ a somewhat different technique for
analysing the environments of AGN by looking at the relative positions
of QSOs with respect to the large scale structure traced out by
clusters and super-cluster structures in the same redshift slices.
They claim that their sample of QSOs follows the large scale
structure, so that QSOs are more likely to be found in the vicinity of
a cluster or in the confluence of a merging cluster system.  This
implies that despite not always residing in rich clusters, QSOs are
nevertheless useful tracers of large scale structure.  Similarly,
\citet{2003MNRAS.346..229B} claim that radio-loud QSOs can be employed 
as efficient tools for detecting high redshift galaxy clusters as they
are often found together in the same fields; although they do warn
that many of the earlier studies are likely to be biased in their
calculations of QSO environmental richness because QSOs are rarely
found directly in the centres of over-densities.

At lower AGN luminosities optically selected Seyfert galaxies seem
much less likely to be found in rich environments.
\citet{1998ApJ...496...93D} analyse a sample of Seyfert
galaxies and find no significant difference between the environmental
richness, or the probability of finding a close companion galaxy,
compared to a matched sample of non-active galaxies.  Although they do
find a difference between the environments of the Seyfert 1 and
Seyfert 2 sub-samples with Seyfert 1s being in poorer environments, an
observation that they cannot explain in terms of the Unified Model of
AGN, which predicts that there should be no difference in the
environment of these two AGN classes. 

In all the above cases the sample sizes have been necessarily small
(typically several tens of QSOs) because of the limitations in
performing large numbers of pointed observations, especially if the
sample is at high redshift (see table 1 of \citet{2001AJ....122...26B}
for a summary of a sample of studies of AGN environments).  The
situation at low redshifts ($z<0.1$), however, is now somewhat
alleviated by large spectroscopic surveys such as the Sloan Digital
Sky Survey (SDSS) or 2df Galaxy Redshift Survey, which include many
thousands of AGN.  Despite this plethora of data different studies
still disagree to some extent on some details of the environments of
AGN.  

\citet{2003ApJ...597..142M} find essentially no change in the
fraction of galaxies with an AGN, across nearly two decades in
environmental density.  Of the nearly 5000 galaxies studied up to
$\sim40~per~cent$ showed some sign of nuclear activity (an upper limit
based on modelling of the lower S/N emission lines; the higher S/N
lines allowed $\sim20~per~cent$ to be unambiguously classified as
AGN), the fraction remaining constant with density.  On the other
hand, star-forming galaxies are found in much greater abundance in
rarefied environments -- the so called SFR-density relation.  Passive
galaxies, of course, are found in greater abundances in denser
environments.  Such a high, and constant, fraction of galaxies
containing an AGN rather suggests that the fuelling mechanism for
these lower luminosity objects (mostly LINERS, the most common and
lowest luminosity AGN class) is a frequent occurrence, and common to a
large range of environments.  Major mergers therefore seem highly
unlikely as a common fuelling mechanism, as do any other cluster
related mechanisms.

\citet{2004MNRAS.353..713K} also use the SDSS data to study the AGN
fraction as a function of environmental density.  They find a somewhat
different result from \citet{2003ApJ...597..142M} in that twice as
many galaxies host AGN in low density environments as in high, a trend
they attribute to the fact that AGN and star-formation are related in
some way.  However, their classification of AGN differs from that of
\citet{2003ApJ...597..142M}, which is possibly the cause of the
difference.  \citet{2004MNRAS.353..713K} only study AGN with O[III]
luminosities $>10^{7}~L_{\odot}$ (total fraction $\sim0.1$), whereas
\citet{2003ApJ...597..142M} study AGN with a wider range of O[III]
luminosities, resulting in the higher overall AGN fraction.  Whereas
AGN with high O[III] luminosities show an environmental dependence,
those with lower O[III] luminosities do not, so in-fact the two
results are not in contradiction.

\citet{2004ApJ...610L..85W} extend the work of
\citet{2003ApJ...597..142M} to a larger sample size and find much the
same AGN fraction ($18~per~cent$).  They study the two-point
correlation function of AGN and compare it to all galaxies, finding no
significant difference between the two, suggesting that AGN follow the
distribution of the normal galaxy population, and are thus unbiased
tracers of mass in the Universe.

\subsection{X-Ray Emission as a Tracer of AGN}
X-ray surveys are very observationally efficient at finding AGN over a
wide range in redshift.  Hard X-ray luminosity in particular is a
highly unbiased measure of AGN power, as the only thing that is being
probed is the accretion rate of the SBH itself; the details of the
exact AGN type and viewing angle are unimportant due to the
penetrating power of hard X-rays.  The narrow [OIII] emission line is
also thought to be an unbiased tracer of AGN because it originates
from beyond the obscuring torus, and is commonly used as a measure of
activity in low-z AGN.  The fairly tight correlation between hard
X-ray and [OIII] luminosity \citep{1999AJ....118.1169X} indicates that
the same physical process is likely to be responsible for both, namely
the accreting SBH.  However, at higher redshifts the [OIII] line
becomes harder to detect in weak AGN as more of the galaxy light falls
into the slit or fibre aperture, washing out the nuclear light.  This
problem does not affect the X-ray emission from AGN, however.

Hard X-ray emission is an excellent tracer of AGN activity because it
is difficult for anything other than a SBH to generate X-ray
luminosities in excess of $10^{42}$ erg s$^{-1}$ ($2-10$~keV).  Hard
X-rays are also affected far less by intrinsic absorption than soft
X-rays and can penetrate large column densities of intervening neutral
hydrogen (up to $\sim 10^{23}$~cm$^{-2}$) that would essentially
completely absorb photons of energy less than $2$ keV.  Of course,
nothing is perfect and for extremely high column densities, resulting
in Compton thick obscuration ($N_{H}\sim 1.5\times
10^{24}$~cm$^{-2}$), even hard X-rays are absorbed.  But for the
purposes of this study we shall ignore Compton thick AGN, with the
assumption that they constitute a relatively small fraction of the
total population (see \citet{2003ApJ...598..886U} for a discussion of
the Compton thick contribution).

\section{Data}
\subsection{X-ray Data}
The details of the {\em XMM\/} data, its reduction and the source
detection algorithm applied to it are described in both papers 1 and
2.  Briefly, the {\em XMM\/} exposures are of $\sim 50$ ks duration,
reaching a $2-10$ keV X-ray flux completeness limit of $\sim 6 \times
10^{-15}$~erg~cm$^{-2}$~s$^{-1}$.

The X-ray point sources were identified with optical objects from the
CFDF catalogues as described in detail in paper 2.  Photometric
redshifts were determined from the multi-band photometry of the
reliably identified AGN.  These AGN form the basis for this present
work.

\subsection{Optical Data}
The CFDF catalogues were derived from a campaign to image one square
degree to $I_{AB}(3\sigma,3\arcsec)\sim25.5$, with comparable depths
in $U$, $B$ and $V$.  The survey area was split into 4 sub-surveys of
$30\arcmin \times 30\arcmin$, two of which were used for the
identification of the X-ray sources in paper 2 (the CFRS 3 and 14-h
fields).  The CFDF data were taken with the {\em Canada-France-Hawaii
Telescope\/} using the UH8K mosaic camera in $B$, $V$ and $I$, with
$U$ data supplied by either the CTIO (3-h field) or the KPNO (14-h
field).  Total exposure times were typically $\sim5$ hours for $B$,
$V$ and $I$, and $\sim10$ hours for $U$.  The lengthy data reduction
process is described in detail in \citep{2001A&A...376..756M}.

\section{Selection of AGN Sample}
\label{AGN_sample}
The selection of the AGN sample requires careful consideration, in
order to avoid uncertainties leading from degeneracies in redshift and
X-ray luminosity, for example.  In any flux limited survey, such as
our {\em XMM\/} surveys, an inevitable correlation arises between
redshift and X-ray luminosity.  Therefore, if we wish to study, for
example, potential correlations between clustering amplitude and X-ray
luminosity, or between clustering amplitude and redshift, then we must
select a sample accordingly so that the trends associated with one
effect are not confused with those caused by the other.

To be able to reduce the error introduced by uncertain background and
foreground number counts, as well as reducing uncertainties in X-ray
luminosity, we require the best possible photometric redshift
estimates for both the AGN and the surrounding field galaxies.
Therefore, we restrict this analysis to only the 14-h field, which has
more accurate photometric redshifts than the 3-h field.  In this work
we use photometric redshifts derived from a slightly improved version
of BPZ \citep{2000ApJ...536..571B}, compared to that used in paper 2;
this version allows the photometric zero-points of the galaxy
catalogue to be adjusted, leading to an improvement in the accuracy of
the photometric redshifts.  After fitting a galaxy template and
redshift to each input galaxy, the code then compares the input
colours with those of the fitted templates for each galaxy.  Any
systematic difference between the input and template colours, for the
whole catalogue, may indicate a systematic photometry error in the
input catalogue, which can be accounted for before re-running the code
by applying a global photometric zero-point offset.  After several
iterations the number of galaxies with reliable redshifts is
increased, with far fewer catastrophic redshift errors.

Adjusting photometric zero-points greatly improves the accuracy and
reliability of the photometric redshifts but we can go one step
further.  To ensure that we only consider galaxies with good redshifts
we construct a reduced CFDF catalogue containing only those galaxies
with a high reliability measure, $P_{\Delta z}>0.9$
\citep{2000ApJ...536..571B}.  Briefly, this quantity represents the
peakedness of the redshift probability function for a particular galaxy
template fit.  It is the integration of the probability function around
the best fit redshift out to limits defined in the code as $\Delta z =
0.2\times(1+z)$.  If the probability function has a well defined peak,
that lies entirely within the integration range, then $P_{\Delta
z}=1.0$, whereas a function that is very broad, or multi-moded, will
have a low $P_{\Delta z}$.  Therefore, $P_{\Delta z}$ gives a measure
of the reliability of a particular photometric redshift estimate.

Using the above criterion further reduces the number of catastrophic
redshift errors in the optical catalogue, by retaining only those
galaxies with a sharp, single-moded redshift probability functions in
the catalogue.  This criterion is also used for the AGN selection,
along with the other criteria described below.

\begin{figure}
\begin{center}
\epsfig{file=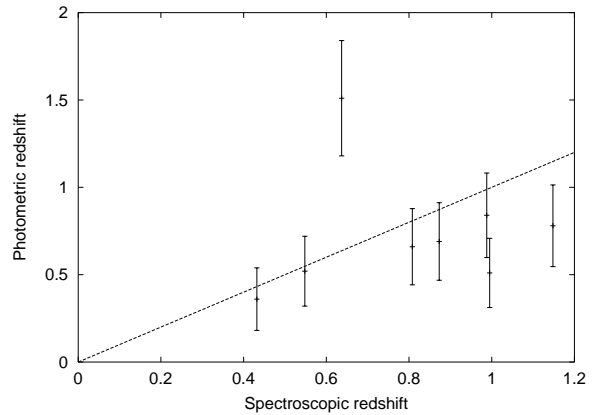,width=0.45\textwidth}
\caption[]{Photometric vs. spectroscopic redshifts for AGN in the 14-h
CFRS field.  Only a handful of AGN in this field have
spectroscopically measured redshifts, hence the need for photometric
redshifts for the remainder.  AGN with unreliable photometric
redshifts ($P_{\Delta z}<0.9$), or with stellar light profiles, have
been removed from this figure leaving only those that we are willing
to consider for selection in the final sample.  The error bars are the
$95~per~cent$ confidence limits on the photometric redshifts, as given
by BPZ.  The dashed line is the expected 1:1 correspondence for
perfect photometric redshifts.}
\label{AGN_z} 
\end{center}
\end{figure}

The photometric redshifts are most accurate for $z<0.6$ (determined
from the whole CFDF catalogue), so to maximise the number of sources
in our AGN sample (because the redshift distribution peaks at $z\sim
0.7$ (paper 2)), while maintaining a narrow enough range to minimise
redshift/luminosity correlations, we select sources in the range
$0.4\leq z \leq0.6$.  For all galaxies with both photometric and
spectroscopic redshifts the rms difference is $\delta z = 0.1$, over
this redshift range.

Figure~\ref{AGN_z} shows the photometric vs. spectroscopic redshifts
for all the non-stellar AGN in the 14-h CFRS field that have a CFRS
measured redshift, and that also have photometric redshifts with a
high reliability measure ($P_{\Delta z}>0.9$).  The $95~per~cent$
confidence limits on the photometric redshifts are shown by the error
bars.  By selecting only AGN with non-stellar light profiles we ensure
that the photometric redshift estimates are not overly affected by
contamination from nuclear light.  \citet{2002ApJ...581..155G}
demonstrate that for X-ray selected AGN BPZ is a reliable way of
obtaining photometric redshifts, as long as the galaxies themselves
are not quasar dominated, so we are confident that our method is
robust.

\begin{figure}
\begin{center}
\epsfig{file=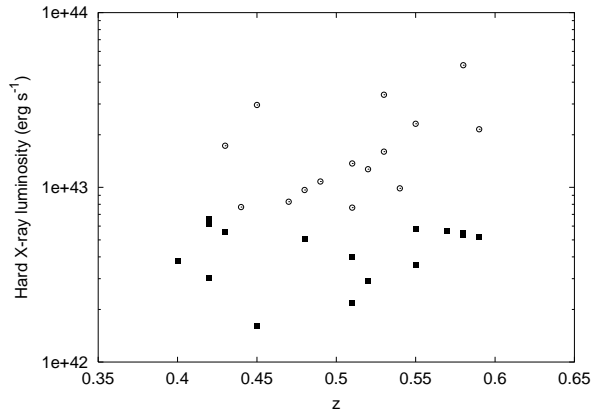,width=0.45\textwidth}
\caption[Luminosity vs. redshift for the 31 AGN in the environment study
sample]{Rest frame $2-10$ keV Luminosity vs. redshift for the 31 AGN
in our environment sample.  The full sample is split into two
sub-samples divided by luminosity: Filled squares, low luminosity (16
objects); open circles, high luminosity (15 objects).}
\label{sample_selection} 
\end{center}
\end{figure}

Figure~\ref{sample_selection} shows the final sample of 31 AGN plotted
with hard X-ray luminosity (calculated assuming a photon index of 1.7
for the K-correction) versus redshift.  We split the full sample into
two sub-samples, based on X-ray luminosity, to test for any
redshift/luminosity correlation; 16 sources are in the low luminosity
sample, 15 are in the high luminosity sample.  Both samples have a
median redshift of 0.51, and a Kolmogorov-Smirnov test shows no
evidence for a difference in the redshift distributions.  Figure 11 of
\citet{2003ApJ...598..886U} demonstrates that for the redshift range
we consider here we are analysing AGN that populate the break in the
hard X-ray luminosity function.  This is important because the break
in any luminosity function with a shallow faint end slope ($\alpha<1$)
constitutes the peak in luminosity-density.  Therefore, sources near
the break effectively contribute more to the luminosity-density of the
population than either lower or higher luminosity sources.  In a
sense, they represent the `average' sources in a population.  Combine
this with the fact that the number-density of AGN also peaks near to
the redshift range we are considering ($z\sim0.7$), and we are clearly
studying an important epoch for accretion onto SBHs.

All but one of the 31 sources in this sample lie above the
$log(f_{X}/f_{opt}) = -1$ line in figure 3 of paper 2 (source 14.144
lies just below), confirming that they are highly likely to be AGN
rather than starburst galaxies.  Starbursts also typically have upper
limits on their hard X-ray luminosities of $\sim10^{41}$ erg s$^{-1}$,
safely below the lower limit for our AGN sample.

\begin{figure*}
\begin{center}
\epsfig{file=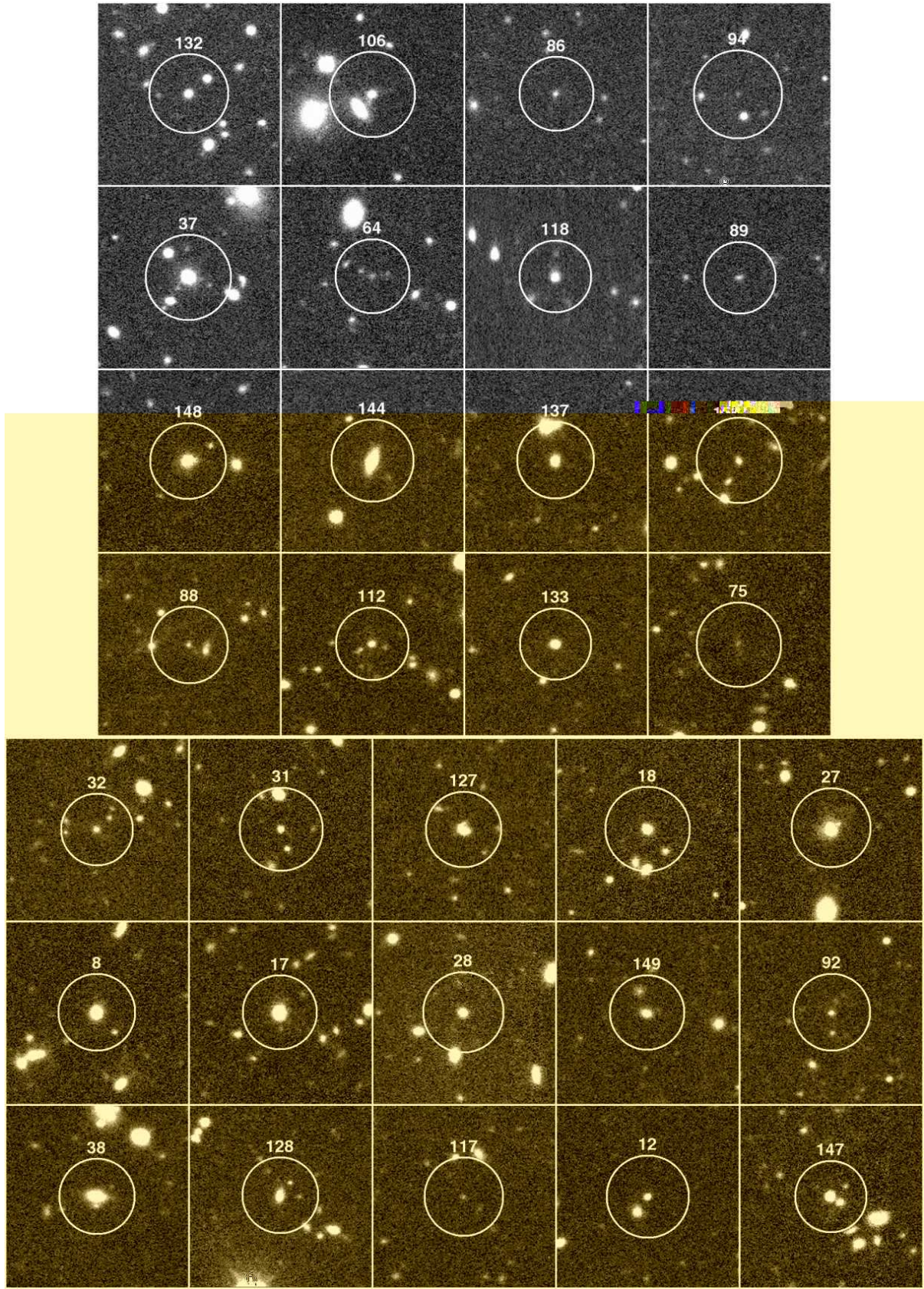,width=0.9\textwidth}
\caption[]{$I$ band thumbnails of the AGN sample described in this
work.  Each image is $\sim 40\arcsec$ square and the circles are 50
kpc physical radius, at the redshift of each AGN.  The low luminosity
sub-sample is at the top, the high luminosity sub-sample is at the
bottom.}
\label{thumbnails} 
\end{center}
\end{figure*}

Figure~\ref{thumbnails} shows thumbnail $I$ band images centred on
each AGN in the sample, labelled as in table~\ref{AGN_clustering}.
The circles that are also displayed all have a physical radius of 50
kpc, while the thumbnail images themselves are all $\sim 40\arcsec$
square.  None of the AGN in this sample has a stellar light profile,
so all are unambiguously extended, as measured by the stellarity
parameter in the SExtractor output catalogue.  Therefore, we can be
fairly confident that the photometric redshifts for this sample are
reliable.  It is also clear from this figure that none of the AGN show
obvious signs of interactions or major distortions arising from recent
mergers.

\section{Calculation of $B_{gq}$}
The clustering amplitude of galaxies around a point of interest gives
a good indication as to the environmental density at that point.  The
quantity $B_{gq}$ is one of the more common measures of the clustering
amplitude and has been used in many of the previous studies into the
environments of quasars
\citep[e.g.][]{2000MNRAS.316..267W,2001MNRAS.321..515M,
2003MNRAS.346..229B}.  We follow the same procedure here, although with
one addition that improves the reliability of the measurements --
through the use of photometric redshift estimates.  

To summarise: the number of galaxies within 0.5 Mpc and $dz\leq 0.1$
(found to be the optimum $dz$ for enhancing the contrast of
over-densities against the background population) of each AGN are
counted (discounting the AGN itself) and compared to the number
expected for the background, as calculated from the total number of
galaxies in the same redshift range for the whole CFDF catalogue
(increasing the value of $dz$ so that the whole input catalogue is
searched has a negligible effect on the results in
section~\ref{results} - save for massively increasing the size of the
error bars due to the higher background count - showing that the
analysis is robust).  What we aim to measure with this is the 2-point
angular correlation function of the galaxies in the vicinity of the
AGN.  We assume that it takes the form:
$$W_{gq}(\theta)=A_{gq}\theta^{1-\gamma}$$ and that $\gamma=1.77$, the
canonical value for the field galaxy population
\citep{1977ApJ...217..385G}.  By integrating $W(\theta)$ out to a radius
$\theta$ the following expression is obtained:
$$A_{gq}=\left[\frac{N_{tot}-N_{b}}{N_{b}}\right]
\left(\frac{3-\gamma}{2}\right)\theta^{\gamma-1}$$ where $N_{tot}$ is
the total number of galaxies within the circle of radius $\theta$ and
$N_{b}$ is the number of background galaxies expected to be found
within the same circle.  At different redshifts the value of $\theta$
is different so as to keep the same physical 0.5 Mpc counting radius
for all the AGN.

The value $A_{gq}$ is the angular clustering amplitude between the AGN
and the surrounding galaxies (equivalent to $\theta_{0}^{\gamma-1}$)
but what we are really after is the {\em spatial\/} clustering
amplitude $B_{gq}$ ($=r_{0}^{\gamma}$), which gives the strength of
the true 3-dimensional 2-point correlation function:
$$\xi_{gq}(r)=B_{gq}r^{-\gamma}.$$ We convert from $A_{gq}$ to
$B_{gq}$ in the same way as \citet{2000MNRAS.316..267W} by using the
following relation:
$$B_{gq}=\frac{N_{g}A_{gq}}{\Phi(m_{lim},z)I_{\gamma}}d_{\theta}^{\gamma-3}$$
where $N_{g}$ is the mean surface density of galaxies per steradian,
$d_{\theta}$ is the angular diameter distance to the AGN and
$I_{\gamma}=3.78$ is an integration constant.  The final quantity,
$\Phi(m_{lim},z)$ is the integrated luminosity function (LF) of
galaxies at the redshift of the AGN, down to some limiting magnitude.
$$\Phi(m_{lim},z)=\int_{L(m_{lim},z)}^{\infty}\phi(L)dL.$$ The
detailed derivation of the conversion from $A_{gq}$ to $B_{gq}$ can be
found in \citet{1979MNRAS.189..433L}.

The error in the clustering amplitude is given by: $$\frac{\Delta
A_{cg}}{A_{cg}}= \frac{\Delta B_{cg}}{B_{cg}}=
\frac{[(N_{tot}-N_{b})+1.3^{2}N_{b}]^{1/2}}{N_{tot}-N_{b}}$$
\citep{1999AJ....117.1985Y}, the $1.3^{2}$ factor coming from
deviations of the field galaxy population from a true Poisson
distribution.

Only galaxies with $I_{AB}<23$ are counted in this study.  The reason
for this (aside from ensuring the most reliable redshifts possible)
is that a compromise must be reached between counting galaxies to too
bright a limit, resulting in low counting statistics, and counting to
too faint a limit, which causes large uncertainties resulting from a
high background count.  A suitable range of $M^*+1$ to $M^*+3$ has
been suggested by \citet{1999AJ....117.1985Y} to optimise the
calculation of $B_{gq}$, but we can afford to go slightly deeper
because we use photometric redshift cuts to improve the contrast of
the AGN regions against the background counts.  Using the Schechter
luminosity function from \citet{2003MNRAS.346..229B}
($M_{I}^{*}=-22.65$, $\alpha=-0.89$ and $\phi^{*}=0.0052$), we reach
$M^*+3.9$ at $z=0.4$ and $M^*+2.8$ at $z=0.6$ using a limit of
$I_{AB}<23$ (for a Sbc galaxy template).  It is prudent to note here
that varying the magnitude limit by $\pm 1$ does not appreciably
change the results, which suggests that the shape of the assumed LF is
indeed suitable for this analysis.  Choosing a limiting magnitude of
$I_{AB}<23$ also reduces the effect of incompleteness in the reduced,
good photometric redshift, CFDF catalogue (see section~\ref{correction}).

\subsection{Control Sample}
\label{control}
A big advantage that the CFDF catalogue has over other similar studies
of AGN environments is that it is a contiguous patch of sky, with many
field galaxies from which to get a reliable estimate of background
galaxy counts.  A further advantage is afforded by the availability of
a large number of galaxies that can be used as a control sample
against which the AGN sample can be compared.  For this study we use
as the control sample all galaxies in a $15\arcmin\times 15\arcmin$
square in the centre of the CFDF map (to avoid edge effects), in the
redshift range $z=[0.4:0.6]$ and with magnitudes $I_{AB}<23$.  We
calculate $B_{gg}$ for the resulting 820 galaxies in exactly the same
way as we calculate $B_{gq}$ for the AGN sample.  

However, the magnitude distribution of this control sample is
different to that of the AGN sample, with a higher proportion of faint
galaxies.  Therefore, a second control sample was extracted from the
first so that it more closely matched the $I_{AB}$ distribution of the
AGN.  To do this we randomly removed fainter galaxies from the
original control sample until it resembled the AGN distribution, with
a much higher matching probability.  For the rest of the discussion
this reduced sample of 297 galaxies will be referred to as the `well
matched control sample'.

\subsection{Correcting for Incompleteness}
\label{correction}
\begin{figure}
\begin{center}
\epsfig{file=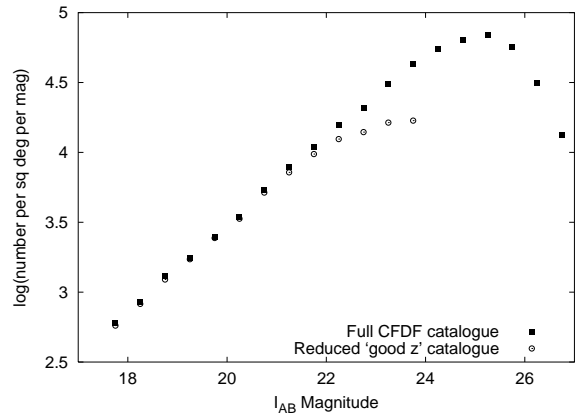,width=0.45\textwidth}
\caption[Number counts of the full and reduced CFDF
catalogue]{Comparison of number counts, in half magnitude bins,
between the full 14-h CFDF catalogue and the reduced sample (shown for
$I_{AB}<24$) containing galaxies with good photometric redshifts.  The
difference between the two catalogues shows the level of correction
required by the reduced sample to account for its incompleteness.}
\label{completeness} 
\end{center}
\end{figure}

Because we perform the above analysis on only galaxies with good
photometric redshifts ($P_{\Delta z}>0.9$) the results will be
affected by a degree of incompleteness in the catalogue.  This effect
is illustrated in figure~\ref{completeness}.  At bright magnitudes
essentially all the galaxies have reliable photometric redshift
estimates.  At the fainter limit, photometric errors cause many of the
galaxies to have unreliable redshift estimates, and so these are lost
from the reduced catalogue; at $I_{AB}=23$ the full CFDF catalogue,
which is still complete (and remains so to at least $I_{AB}=25$),
contains $\sim 50~per~cent$ more galaxies than the reduced sample.
Therefore, to ensure that the results in this work are not biased it
is necessary to correct the reduced sample by the incompleteness
factor at any given magnitude.  To do this we simply multiply the
number of galaxies of a given magnitude by the required factor to
bring the number up to that expected from the full catalogue.  At most
this difference is a factor of 1.5 for magnitudes in the range
$I_{AB}=[22.5:23]$.  Because the counts around the AGN and the
background counts are corrected in the same way, this correction
should not affect our conclusions.  In fact the same is true for the
exact details of the assumed LF; as long as the same LF is used for
the control samples as for the AGN sample then the absolute measure of
clustering amplitude is unimportant -- it is the {\em relative\/}
clustering amplitudes that reveal the important facts.

\section{Results}
\label{results}
\begin{figure*}
\begin{center}
\epsfig{file=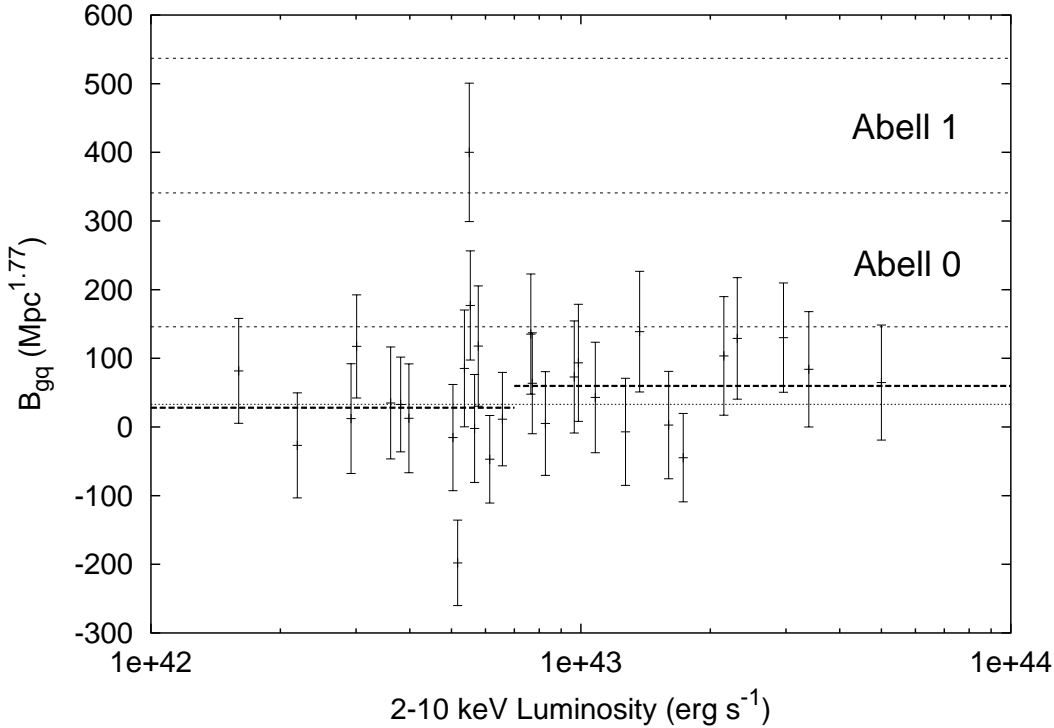,width=0.8\textwidth}
\caption[Clustering amplitude $B_{gq}$ of galaxies around the 14-h AGN
environment sample.]{Clustering amplitude of galaxies in the vicinity
of 31 AGN in the 14-h field (see tables~\ref{AGN_clustering}
and~\ref{clustering_results}).  The thin dotted line at $B_{gq}\sim35$
Mpc$^{1.77}$ is the mean value for 820 field galaxies, drawn from the
same redshift range as the AGN sample and analysed in an identical
fashion.  The thicker, short-dashed lines show the mean values for the
two sub-samples of AGN: left, low luminosity; right, high luminosity.
Regions corresponding to Abell richness classes 0 and 1 are delineated
by the dotted lines at 146, 341 and 537 Mpc$^{1.77}$; class 2 lies
above the top line (values taken from \citet{2001MNRAS.321..515M} and
re-scaled to match our chosen cosmology).}
\label{Bgq} 
\end{center}
\end{figure*}

Table~\ref{AGN_clustering} shows the results for both $A_{gq}$ and
$B_{gq}$ for the AGN sample.  Figure~\ref{Bgq} shows these same
results plotted with $B_{gq}$ vs. hard X-ray luminosity.
Additionally, the mean values for the two AGN sub-samples and the
field galaxy sample are also plotted and are tabulated in
table~\ref{clustering_results}

\begin{table*}
\begin{center}
\caption[]{Results of the clustering amplitude for the 31 AGN in the
range $0.4\leq z\leq0.6$.  The coordinates are J2000 and are those of
the optical identifications for the X-ray sources.  The redshifts are
those given by the BPZ code (see text for details).  The optical
magnitudes are AB and are measured inside a $3\arcsec$ diameter
aperture.  Luminosity is measured in the rest frame hard X-ray band
($2-10$ keV) and has units of erg s$^{-1}$.  Low and high luminosity
sub-samples are divided by the horizontal line.}
\label{AGN_clustering}
\vspace{0.5cm}
\begin{tabular}{lcccccccrrrr} \hline
\multicolumn{1}{c}{XID}&
\multicolumn{1}{c}{R.A.}&
\multicolumn{1}{c}{Dec.}&
\multicolumn{1}{c}{z}&
\multicolumn{1}{c}{$U$}&
\multicolumn{1}{c}{$B$}&
\multicolumn{1}{c}{$V$}&
\multicolumn{1}{c}{$I$}&
\multicolumn{1}{c}{Luminosity}&
\multicolumn{1}{c}{$A_{gq}$}&
\multicolumn{1}{c}{$B_{gq}$ (Mpc$^{1.77}$)}&
\multicolumn{1}{c}{$\Delta B$}\\ \hline
132 & 214.0588 & 52.32756 & 0.48 & 23.23 & 22.98 & 22.41 & 21.62 & 5.50E+42 & 1.94E-03 & 400.0 & 100.8\\
106 & 214.0621 & 52.51536 & 0.42 & 23.40 & 22.83 & 22.04 & 21.15 & 6.57E+42 & 6.91E-05 & 11.4 & 68.0\\
86  & 214.0640 & 52.47858 & 0.55 & 25.26 & 25.49 & 24.29 & 22.76 & 5.17E+42 & -9.65E-04 & -197.9 & 62.3\\
94  & 214.0697 & 52.44880 & 0.40 & 25.46 & 25.88 & 25.06 & 24.41 & 3.81E+42 & 2.00E-04 & 32.8 & 69.0\\
37  & 214.1584 & 52.38561 & 0.42 & 24.04 & 22.72 & 21.45 & 19.86 & 3.01E+42 & 7.13E-04 & 117.4 & 75.1\\
64  & 214.1682 & 52.37271 & 0.55 & 24.58 & 24.98 & 24.27 & 23.34 & 2.92E+42 & 5.94E-05 & 12.2 & 80.0\\
118 & 214.2061 & 52.28159 & 0.58 & 23.48 & 23.36 & 22.67 & 21.47 & 5.77E+42 & 5.95E-04 & 117.9 & 87.7\\
89  & 214.2162 & 52.45008 & 0.58 & 26.70 & 25.24 & 24.52 & 22.95 & 5.66E+42 & -1.09E-05 & -2.2 & 78.6\\
148 & 214.2870 & 52.45245 & 0.52 & 24.55 & 23.50 & 22.17 & 20.45 & 2.19E+42 & -1.29E-04 & -26.7 & 76.5\\
144 & 214.3141 & 52.32068 & 0.45 & 23.66 & 22.85 & 21.68 & 20.17 & 1.60E+42 & 4.59E-04 & 81.7 & 76.4\\
137 & 214.3911 & 52.34428 & 0.51 & 23.39 & 22.92 & 22.08 & 20.89 & 5.04E+42 & -7.35E-05 & -15.4 & 77.3\\
153 & 214.4041 & 52.59928 & 0.42 & 23.97 & 23.57 & 22.97 & 22.29 & 6.14E+42 & -2.86E-04 & -47.0 & 63.7\\
88  & 214.4549 & 52.46978 & 0.51 & 24.80 & 24.67 & 24.00 & 23.07 & 3.98E+42 & 6.09E-05 & 12.7 & 79.3\\
112 & 214.4618 & 52.27126 & 0.57 & 25.14 & 24.47 & 23.53 & 21.87 & 3.61E+42 & 1.74E-04 & 35.0 & 81.6\\
133 & 214.4705 & 52.47736 & 0.59 & 23.15 & 22.76 & 22.31 & 21.11 & 5.36E+42 & 4.40E-04 & 85.3 & 85.1\\
75  & 214.4709 & 52.29158 & 0.43 & 25.86 & 25.29 & 24.43 & 23.49 & 5.53E+42 & 1.06E-03 & 176.9 & 79.5\\ \hline
32  & 214.1363 & 52.31716 & 0.58 & 24.14 & 24.40 & 23.51 & 22.29 & 5.00E+43 & 3.27E-04 & 64.8 & 83.8\\
31  & 214.1437 & 52.37738 & 0.44 & 23.82 & 23.46 & 22.64 & 21.78 & 7.71E+42 & 3.70E-04 & 63.6 & 73.5\\
127 & 214.1439 & 52.19802 & 0.53 & 23.48 & 23.15 & 22.29 & 21.10 & 3.39E+43 & 4.10E-04 & 84.0 & 84.1\\
18  & 214.1747 & 52.52856 & 0.43 & 24.44 & 23.42 & 22.25 & 20.64 & 1.73E+43 & -2.68E-04 & -44.7 & 64.4\\
27  & 214.1832 & 52.37185 & 0.48 & 23.56 & 22.65 & 21.74 & 20.11 & 9.65E+42 & 3.53E-04 & 72.8 & 81.6\\
8   & 214.2527 & 52.32180 & 0.51 & 21.88 & 21.62 & 21.13 & 20.27 & 1.37E+43 & 6.65E-04 & 138.9 & 87.8\\
17  & 214.2992 & 52.33660 & 0.55 & 21.93 & 21.67 & 21.00 & 19.89 & 2.31E+43 & 6.29E-04 & 129.0 & 88.4\\
28  & 214.3472 & 52.53151 & 0.47 & 22.67 & 22.22 & 21.85 & 21.12 & 8.27E+42 & 2.55E-05 & 5.1 & 75.5\\
149 & 214.3699 & 52.59755 & 0.54 & 24.25 & 23.51 & 22.75 & 21.39 & 9.87E+42 & 4.56E-04 & 93.5 & 85.3\\
92  & 214.3745 & 52.62106 & 0.51 & 25.64 & 24.67 & 23.92 & 22.52 & 7.65E+42 & 6.47E-04 & 135.3 & 87.6\\
38  & 214.3749 & 52.20760 & 0.53 & 22.39 & 21.90 & 21.19 & 19.98 & 1.60E+43 & 1.34E-05 & 2.8 & 78.3\\
128 & 214.3907 & 52.56351 & 0.52 & 23.30 & 22.84 & 22.19 & 21.03 & 1.27E+43 & -3.43E-05 & -7.1 & 78.0\\
117 & 214.4110 & 52.57036 & 0.49 & 25.30 & 25.25 & 24.66 & 23.88 & 1.08E+43 & 2.06E-04 & 43.0 & 80.5\\
12  & 214.5084 & 52.30996 & 0.45 & 22.25 & 22.51 & 21.99 & 21.80 & 2.96E+43 & 7.31E-04 & 130.1 & 79.6\\
147 & 214.5235 & 52.29270 & 0.59 & 23.62 & 23.09 & 22.26 & 20.74 & 2.15E+43 & 5.33E-04 & 103.5 & 86.4\\ \hline
\end{tabular}
\end{center}
\end{table*}

\begin{table*}
\begin{center}
\caption[Clustering amplitude averages.]{Results for the clustering
amplitude $B_{gq}$ in Mpc$^{1.77}$.}
\label{clustering_results}
\vspace{0.5cm}
\begin{tabular}{lccc} \hline
\multicolumn{1}{l}{Sample}&
\multicolumn{1}{c}{Weighted mean}&
\multicolumn{1}{c}{Median}&
\multicolumn{1}{c}{Straight mean}\\ \hline 
820 control galaxies       & $36.6\pm2.7$  & 44.9 & 53.3\\
297 well matched control   & $42.9\pm4.5$  & 54.5 & 61.0\\
All 31 AGN                 & $42.5\pm14.0$ & 63.6 & 58.3\\
16 low luminosity AGN      & $28.1\pm19.0$ & 22.8 & 49.6\\
15 high luminosity AGN     & $59.7\pm20.7$ & 72.8 & 67.6\\ \hline
\end{tabular}
\end{center}
\end{table*}

Notice how the well matched control sample has a slightly higher
clustering amplitude than the full control sample, due primarily to
the higher proportion of relatively brighter galaxies in the matched
sample.  However, the difference between the environments of the two
control samples is not really significant and does not make any
qualitative difference to the final results.

The first thing that is obvious from these results is that the AGN
sample is not significantly different from that of either control
sample.  To formalise this we perform K-S tests on each of the sample
pairs listed in table~\ref{results_KS}.  As table~\ref{results_KS}
shows the AGN sample is {\em indistinguishable\/} from the field
galaxy population, when either the full control or well matched
control samples are used.  The very slight hint of a difference
between the two AGN samples is also not significant.

\begin{table}
\begin{center}
\caption[K-S tests on the clustering results.]{K-S tests to determine
if the clustering amplitudes for the AGN are drawn from a different
population to the field control samples.  Here $P'$ is the the
probability that the two samples are {\em not\/} drawn from the same
population.}
\label{results_KS}
\vspace{0.5cm}
\begin{tabular}{llc} \hline
\multicolumn{2}{c}{Samples}&
\multicolumn{1}{c}{P'}\\ \hline 
Full control    & All AGN        & 0.297\\
Full control    & Low $L_X$ AGN  & 0.287\\
Full control    & High $L_X$ AGN & 0.708\\
Matched control & All AGN        & 0.573\\
Matched control & Low $L_X$ AGN  & 0.408\\
Matched control & High $L_X$ AGN & 0.588\\
Low $L_X$ AGN   & High $L_X$ AGN & 0.784\\ 
Matched control & Full control   & 0.138\\ \hline
\end{tabular}
\end{center}
\end{table}

\section{Close Companions}
So it seems from the clustering amplitude analysis that the Mpc scale
environments of moderate luminosity AGN are essentially the same as
those of non-active galaxies.  In this section we investigate the
possibility that tidal interactions with nearby galaxies are important 
as fuelling mechanisms for these AGN.  Again, we compare the AGN sample 
to the two control samples described in section~\ref{control}.

For this analysis we extend the magnitude range of the search to
$I_{AB}<24$, so one magnitude fainter than the clustering amplitude
analysis i.e. $M^*+4.9$ at $z=0.4$ and $M^*+3.8$ at $z=0.6$ (similar
to the SMC and LMC).  We also neglect the effects of completion here
because we are making a direct comparison between samples that should
be affected by completeness in an identical way, and therefore an
absolute measure is unnecessary.

Table~\ref{companions} shows the number of galaxies found within a
given radius of galaxies in the AGN and control samples.  It is clear
that the environments of AGN host galaxies are very similar to those
of inactive galaxies on all the scales investigated here.  

\begin{figure}
\begin{center}
 \subfigure[\label{30kpc_hist}]{\epsfig{file=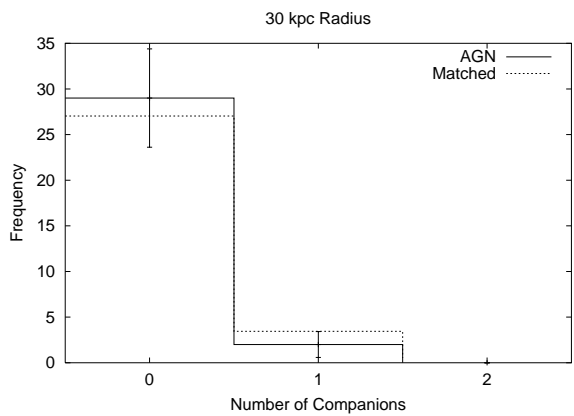,width=0.45\textwidth}}
 \subfigure[\label{50kpc_hist}]{\epsfig{file=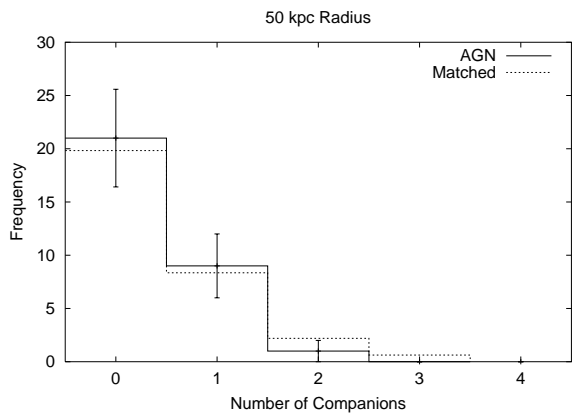,width=0.45\textwidth}}
 \caption[Histograms of the number of companion galaxies.]{Histograms
 of the number of companion galaxies found within 30 kpc
 (\ref{30kpc_hist}) and 50 kpc (\ref{50kpc_hist}) of the galaxies in
 the AGN sample.  The well matched control sample (scaled to match the
 AGN sample) is shown for comparison as the dashed histograms.  Error
 bars on the AGN sample are Poisson uncertainties.  An AGN host is no
 more or less likely to have a close companion than an inactive
 galaxy.}
\label{companions_histograms}
\end{center}
\end{figure}

Figure~\ref{companions_histograms} shows the distribution of the
frequency of different numbers of companion galaxies on the two
smallest scales.  On these scales the number of companion galaxies is
small but the AGN sample has essentially the same distribution as that
of the well matched control sample.  The same is true of the larger
scales.

\begin{table*}
\begin{center}
\caption[Mean number of companion galaxies.]{Mean number of companion
galaxies for the various samples, counted within different radii ($dz\leq
0.1$, $I_{AB}<24$).  Errors are Poisson uncertainties.  The AGN sample
and control samples are all remarkably similar from 30 kpc to 0.5
Mpc}
\label{companions}
\vspace{0.5cm}
\begin{tabular}{lccccc} \hline
\multicolumn{1}{c}{}&
\multicolumn{5}{c}{Counting Radius (kpc)}\\ \hline
\multicolumn{1}{l}{Sample}&
\multicolumn{1}{c}{30}&
\multicolumn{1}{c}{50}&
\multicolumn{1}{c}{100}&
\multicolumn{1}{c}{250}&
\multicolumn{1}{c}{500}\\ \hline 
Full control & $0.14 \pm 0.01$ & $0.43 \pm 0.02$ & $1.61 \pm 0.04$ & $9.85 \pm 0.11$ & $38.3 \pm 0.2$ \\
Well matched & $0.14 \pm 0.02$ & $0.47 \pm 0.04$ & $1.79 \pm 0.08$ & $10.2 \pm 0.19$ & $40.2 \pm 0.4$ \\
All AGN      & $0.07 \pm 0.05$ & $0.35 \pm 0.11$ & $1.45 \pm 0.22$ & $9.65 \pm 0.56$ & $39.5 \pm 1.1$ \\
Low AGN      & $0.00 \pm 0.00$ & $0.19 \pm 0.11$ & $1.44 \pm 0.30$ & $8.56 \pm 0.73$ & $38.0 \pm 1.5$ \\
High AGN     & $0.13 \pm 0.09$ & $0.53 \pm 0.19$ & $1.47 \pm 0.31$ & $10.8 \pm 0.85$ & $41.1 \pm 1.7$ \\ \hline
\end{tabular}
\end{center}
\end{table*}

Figure~\ref{thumbnails} shows the local environments of the individual
AGN for comparison.

So this appears to support the conclusions of
\citet{1998ApJ...496...93D}: the likelihood of finding a 
companion galaxy with $R<-17.5$ within 50 kpc of a Seyfert galaxy is
not statistically different from that for an inactive galaxy.  AGN
with sub-quasar luminosities have essentially identical environments --
from 30 kpc up to 0.5 Mpc -- to those of `normal', inactive galaxies.

\section{Discussion}
\subsection{Implications for AGN Fuelling Mechanisms}
The typical environments of our AGN sample are no different to those of
inactive galaxies in general.  Aside from one example, all the AGN are
found in sub-cluster richness regions, in contrast to the studies of
high luminosity AGN, such as radio-loud QSOs.  Therefore, any fuelling
mechanism requiring the presence or proximity of a rich cluster is
unlikely to be important for fuelling lower luminosity AGN.
Harassment \citep{1998ApJ...495..152L}, for example, may be an
efficient method of transporting gas into the central 500 pc of a gas
rich galaxy but it requires that galaxy to be situated in the
outskirts of a rich cluster.  This is clearly not the case for the
vast majority of AGN, which exist far from the influence of clusters
and, in a purely numerical sense, harassment is simply not capable of
causing up to $40\%$ of all galaxies to be active at any given time
\citep{2003ApJ...597..142M}.

The similarity in the number of close companions between the active and
non-active galaxies suggests that major mergers are also unlikely as a
fuelling mechanism for the lowest luminosity AGN, especially given the
high fraction of galaxies that contain one.  We haven't performed any
asymmetry analysis on the host galaxies of our AGN sample (to look for
signs of recent major mergers or tidal interactions), but
\citet{2003ApJ...595..685G} find that the AGN in the CDF-S (roughly
comparable to the luminosities of the AGN in our sample) are no more
asymmetric than the field galaxy population, suggesting that they are
not under the influence of a recent merger or tidal interaction.
\citet{2003ApJ...595..685G} also investigate the near neighbour
frequency of both the active and non-active galaxies in that field and
essentially confirm our result, that there is no significant difference
between the two populations.  \citet{1998ApJ...496...93D} also confirm
both these results for Seyfert galaxies, which typically have hard
X-ray luminosities in the range of our AGN sample.

But could we be missing really close companions in our near neighbour
analysis?  Our close companion search is limited to $I_{AB}<24$ and at
this faint magnitude limit incompleteness accounts for the loss of
over half the galaxies from the reduced CFDF catalogue, which we use
in this analysis.  Conceivably, most of these losses could be the
nearest companions of brighter galaxies.  At close proximity the
photometry of a faint galaxy could be contaminated by light from its
brighter neighbour, leading to an unreliable photometric redshift and
its subsequent rejection from the catalogue.  Therefore, if one sample
in the analysis has a real excess of close, faint companions relative
to another sample, then this difference will be suppressed by the
preferential loss of those close companions from the catalogue.
We believe this possible observational bias {\em is\/} present in
this work but does not make a major impact on the overall conclusions,
as explained below.

The CFDF photometry is measured using a $3\arcsec$ diameter aperture,
which equates to a physical radius of $\sim9$ kpc at $z=0.6$ (roughly
the visible extent of $L^{*}$ galaxies); for the photometry of a
companion galaxy to be significantly affected by the light from
another, brighter galaxy it would have to be less than twice this sort
of distance from it, say 20 kpc to be safe.  The companion probability
for a search radius of 30 kpc is $12.8~per~cent$ (38/297 for the well
matched control sample) and since the galaxy population is clustered
the probability of finding a companion within 20 kpc will be less than
this $12.8~per~cent$ but more than the $6.4~per~cent$ (19/297)
probability obtained by assuming a uniform surface density of
galaxies.  Reducing the counting radius for the well matched control
sample to 20 kpc we find a companion probability of $5.7~per~cent$
(17/297).  So it does appear that we may be preferentially losing some
faint galaxies from the catalogue that are close to brighter galaxies.

However, at this level of probability it would require a much larger
AGN sample to be able to detect a significant deviation from the
control sample (at 20 kpc the companion probability is $3.2\%$ for the
AGN sample, i.e. 1 out of 31), as shown by the error bars in
figure~\ref{companions_histograms}.  So, it is possible that
incompleteness may be responsible for the similarity between the
different samples on the 30 kpc scale, but for the 50 kpc counting
radius and above this effect becomes increasingly less important and
so the observed similarity of the AGN and control samples should be
real.

In addition, the above argument is not applicable to galaxies
undergoing a major merger, in which a companion will tend to be
comparably bright; it is the fainter galaxies that are lost from the
catalogue.  Inspection of figure~\ref{thumbnails} reveals a lack of
companions with similar optical luminosities, so we believe that
incompleteness at faint magnitudes will not affect the conclusion that
major mergers are unimportant as an AGN fuelling mechanism.

So that leaves the leading contender for low luminosity AGN fuelling
as minor mergers.  In much the same way as a major merger or
interaction disrupts the eventual AGN host galaxy, the accretion of a
small satellite galaxy or primordial gas cloud will have the same
effect but on a smaller scale and without the extreme deformation of
the host disc \citep[e.g.][]{1996ApJ...460..121W,1995ApJ...448...41H}.
And since these small objects are very much more numerous than massive
galaxies such minor mergers will be correspondingly much more common
than major ones.  If the accretion of a satellite is onto a gas rich
host then the satellite need only provide the impetus to send the
host's gas in towards the awaiting central engine.  A gas poor
elliptical galaxy, on the other hand, requires the satellite to also
provide the fuel necessary for the nuclear activity.  In the former
case the structure of the host galaxy seems in itself to be an
important factor in determining whether the gas supply is used in a
nuclear starburst or accretion onto a SBH
\citep{1995ApJ...448...41H,1994ApJ...425L..13M}.  If minor mergers are
common occurrences, which they undoubtedly are relative to major ones,
then they are likely to be an important mechanism for the fuelling of
AGN, with the differences between AGN classes being determined by the
properties of the host galaxy.  Unfortunately, catching these events
in the process of happening is observationally challenging, so this
hypothesis is likely to remain unproven for a while yet.

\subsection{Caveats}
One important point to note in this discussion is the limited area
that this study covers.  The survey area of $30\arcmin$ square does
not contain a range of environments as extreme as the richest
clusters.  $20~per~cent$ of the galaxies in the well-matched control
sample reside in environments as rich as Abell $R=0$ clusters, while
$<1~per~cent$ reside in environments as rich as Abell $R=1$.  There
are no environments of greater richness within the redshift range
studied.  Also, there are no extended sources in the X-ray data to
indicate the presence of rich clusters.  However, this does not dilute
the conclusion that the vast majority of moderate X-ray luminous AGN
reside in similar environments to normal galaxies.  Rich clusters are
rare and contain only a small fraction of the total galaxy population.

\section{Future Work}
At present there are no low-z equivalents to this work.  Although the
environments of AGN have been studied in the local Universe
\citep[e.g.][]{1998ApJ...496...93D,2003ApJ...597..142M} the samples
were not compiled using hard X-ray selection.  To properly investigate
the evolution of X-ray selected AGN environments a low-z benchmark is
required, against which higher-z studies, such as this one, can be
compared.  A local Universe study may also be able to detect the
effects of minor merger activity more directly than our high-z study
is able to do.

Extension to higher redshift should also be possible.
Multi-wavelength surveys (combining deep X-ray and multi-band optical
data) are becoming increasingly common now; many of which are also
publicly available.  An almost exact copy of this work could be
applied to the {\em Chandra\/} deep fields for example.  This would
allow the investigation of possible trends associated with X-ray
luminosity (at a fixed redshift) or with redshift (for a fixed
luminosity).

Further sub-division of the X-ray selected AGN into different classes
(based on their best fitting optical templates for example) would help
to determine if the environmental properties of AGN of fixed
luminosity are strongly correlated with the host galaxy properties, as
has been suggested \citep{2003ApJ...597..142M}.

Ideally, future surveys should also be large enough to cover a wider range
of environmental densities than is probed in this current work.

\section*{acknowledgements}
This paper was based on observations obtained with {\em XMM-Newton\/},
an ESA science mission with instruments and contributions directly
funded by ESA Member States and NASA.  The authors would like to thank
the referee for comments that improved the clarity of this paper.

\bibliography{/home/serpens3/berlacie2/spxtjw/papers/bib}

\begin{thebibliography}{}

\bibitem[\protect\citeauthoryear{{Barr}, {Bremer}, {Baker} \& {Lehnert}}{{Barr}
  et~al.}{2003}]{2003MNRAS.346..229B}
{Barr} J.~M.,  {Bremer} M.~N.,  {Baker} J.~C.,    {Lehnert} M.~D.,  2003,
  mnras, 346, 229

\bibitem[\protect\citeauthoryear{{Ben{\'{\i}}tez}}{{Ben{\'{\i}}tez}}{2000}]{20%
00ApJ...536..571B}
{Ben{\'{\i}}tez} N.,  2000, apj, 536, 571

\bibitem[\protect\citeauthoryear{{Brown}, {Boyle} \& {Webster}}{{Brown}
  et~al.}{2001}]{2001AJ....122...26B}
{Brown} M.~J.~I.,  {Boyle} B.~J.,    {Webster} R.~L.,  2001, aj, 122, 26

\bibitem[\protect\citeauthoryear{{De Robertis}, {Yee} \& {Hayhoe}}{{De
  Robertis} et~al.}{1998}]{1998ApJ...496...93D}
{De Robertis} M.~M.,  {Yee} H.~K.~C.,    {Hayhoe} K.,  1998, apj, 496, 93

\bibitem[\protect\citeauthoryear{{Fabian}, {Arnaud}, {Nulsen} \&
  {Mushotzky}}{{Fabian} et~al.}{1986}]{1986ApJ...305....9F}
{Fabian} A.~C.,  {Arnaud} K.~A.,  {Nulsen} P.~E.~J.,    {Mushotzky} R.~F.,
  1986, apj, 305, 9

\bibitem[\protect\citeauthoryear{{Gonzalez} \& {Maccarone}}{{Gonzalez} \&
  {Maccarone}}{2002}]{2002ApJ...581..155G}
{Gonzalez} A.~H.,  {Maccarone} T.~J.,  2002, apj, 581, 155

\bibitem[\protect\citeauthoryear{{Grogin}, {Koekemoer}, {Schreier}, {Bergeron},
  {Giacconi}, {Hasinger}, {Kewley}, {Norman}, {Rosati}, {Tozzi} \&
  {Zirm}}{{Grogin} et~al.}{2003}]{2003ApJ...595..685G}
{Grogin} N.~A.,  {Koekemoer} A.~M.,  {Schreier} E.~J.,  {Bergeron} J.,
  {Giacconi} R.,  {Hasinger} G.,  {Kewley} L.,  {Norman} C.,  {Rosati} P.,
  {Tozzi} P.,    {Zirm} A.,  2003, apj, 595, 685

\bibitem[\protect\citeauthoryear{{Groth} \& {Peebles}}{{Groth} \&
  {Peebles}}{1977}]{1977ApJ...217..385G}
{Groth} E.~J.,  {Peebles} P.~J.~E.,  1977, apj, 217, 385

\bibitem[\protect\citeauthoryear{{Hall}, {Ellingson} \& {Green}}{{Hall}
  et~al.}{1997}]{1997AJ....113.1179H}
{Hall} P.~B.,  {Ellingson} E.,    {Green} R.~F.,  1997, aj, 113, 1179

\bibitem[\protect\citeauthoryear{{Hernquist} \& {Mihos}}{{Hernquist} \&
  {Mihos}}{1995}]{1995ApJ...448...41H}
{Hernquist} L.,  {Mihos} J.~C.,  1995, apj, 448, 41

\bibitem[\protect\citeauthoryear{{Kauffmann} \& {Haehnelt}}{{Kauffmann} \&
  {Haehnelt}}{2000}]{2000MNRAS.311..576K}
{Kauffmann} G.,  {Haehnelt} M.,  2000, mnras, 311, 576

\bibitem[\protect\citeauthoryear{{Kauffmann}, {White}, {Heckman}, {M{\'
  e}nard}, {Brinchmann}, {Charlot}, {Tremonti} \& {Brinkmann}}{{Kauffmann}
  et~al.}{2004}]{2004MNRAS.353..713K}
{Kauffmann} G.,  {White} S.~D.~M.,  {Heckman} T.~M.,  {M{\' e}nard} B.,
  {Brinchmann} J.,  {Charlot} S.,  {Tremonti} C.,    {Brinkmann} J.,  2004,
  mnras, 353, 713

\bibitem[\protect\citeauthoryear{{Lake}, {Katz} \& {Moore}}{{Lake}
  et~al.}{1998}]{1998ApJ...495..152L}
{Lake} G.,  {Katz} N.,    {Moore} B.,  1998, apj, 495, 152

\bibitem[\protect\citeauthoryear{{Longair} \& {Seldner}}{{Longair} \&
  {Seldner}}{1979}]{1979MNRAS.189..433L}
{Longair} M.~S.,  {Seldner} M.,  1979, mnras, 189, 433

\bibitem[\protect\citeauthoryear{{McCracken}, {Le F{\` e}vre}, {Brodwin},
  {Foucaud}, {Lilly}, {Crampton} \& {Mellier}}{{McCracken}
  et~al.}{2001}]{2001A&A...376..756M}
{McCracken} H.~J.,  {Le F{\` e}vre} O.,  {Brodwin} M.,  {Foucaud} S.,  {Lilly}
  S.~J.,  {Crampton} D.,    {Mellier} Y.,  2001, aap, 376, 756

\bibitem[\protect\citeauthoryear{{McLure} \& {Dunlop}}{{McLure} \&
  {Dunlop}}{2001}]{2001MNRAS.321..515M}
{McLure} R.~J.,  {Dunlop} J.~S.,  2001, mnras, 321, 515

\bibitem[\protect\citeauthoryear{{McLure}, {Kukula}, {Dunlop}, {Baum}, {O'Dea}
  \& {Hughes}}{{McLure} et~al.}{1999}]{1999MNRAS.308..377M}
{McLure} R.~J.,  {Kukula} M.~J.,  {Dunlop} J.~S.,  {Baum} S.~A.,  {O'Dea}
  C.~P.,    {Hughes} D.~H.,  1999, mnras, 308, 377

\bibitem[\protect\citeauthoryear{{Mihos} \& {Hernquist}}{{Mihos} \&
  {Hernquist}}{1994}]{1994ApJ...425L..13M}
{Mihos} J.~C.,  {Hernquist} L.,  1994, apjl, 425, L13

\bibitem[\protect\citeauthoryear{{Miller}, {Nichol}, {G{\' o}mez}, {Hopkins} \&
  {Bernardi}}{{Miller} et~al.}{2003}]{2003ApJ...597..142M}
{Miller} C.~J.,  {Nichol} R.~C.,  {G{\' o}mez} P.~L.,  {Hopkins} A.~M.,
  {Bernardi} M.,  2003, apj, 597, 142

\bibitem[\protect\citeauthoryear{{Moore}, {Katz} \& {Lake}}{{Moore}
  et~al.}{1996}]{1996ApJ...457..455M}
{Moore} B.,  {Katz} N.,    {Lake} G.,  1996, apj, 457, 455

\bibitem[\protect\citeauthoryear{{S{\" o}chting}, {Clowes} \&
  {Campusano}}{{S{\" o}chting} et~al.}{2004}]{2004MNRAS.347.1241S}
{S{\" o}chting} I.~K.,  {Clowes} R.~G.,    {Campusano} L.~E.,  2004, mnras,
  347, 1241

\bibitem[\protect\citeauthoryear{{Smith}, {Boyle} \& {Maddox}}{{Smith}
  et~al.}{2000}]{2000MNRAS.313..252S}
{Smith} R.~J.,  {Boyle} B.~J.,    {Maddox} S.~J.,  2000, mnras, 313, 252

\bibitem[\protect\citeauthoryear{{Ueda}, {Akiyama}, {Ohta} \& {Miyaji}}{{Ueda}
  et~al.}{2003}]{2003ApJ...598..886U}
{Ueda} Y.,  {Akiyama} M.,  {Ohta} K.,    {Miyaji} T.,  2003, apj, 598, 886

\bibitem[\protect\citeauthoryear{{Wake}, {Miller}, {Di Matteo}, {Nichol},
  {Pope}, {Szalay}, {Gray}, {Schneider} \& {York}}{{Wake}
  et~al.}{2004}]{2004ApJ...610L..85W}
{Wake} D.~A.,  {Miller} C.~J.,  {Di Matteo} T.,  {Nichol} R.~C.,  {Pope} A.,
  {Szalay} A.~S.,  {Gray} A.,  {Schneider} D.~P.,    {York} D.~G.,  2004, apjl,
  610, L85

\bibitem[\protect\citeauthoryear{{Walker}, {Mihos} \& {Hernquist}}{{Walker}
  et~al.}{1996}]{1996ApJ...460..121W}
{Walker} I.~R.,  {Mihos} J.~C.,    {Hernquist} L.,  1996, apj, 460, 121

\bibitem[\protect\citeauthoryear{{Waskett}, {Eales}, {Gear}, {McCracken},
  {Brodwin}, {Nandra}, {Laird} \& {Lilly}}{{Waskett}
  et~al.}{2004}]{2004MNRAS.350..785W}
{Waskett} T.~J.,  {Eales} S.~A.,  {Gear} W.~K.,  {McCracken} H.~J.,  {Brodwin}
  M.,  {Nandra} K.,  {Laird} E.~S.,    {Lilly} S.,  2004, mnras, 350, 785

\bibitem[\protect\citeauthoryear{{Waskett}, {Eales}, {Gear}, {Puchnarewicz},
  {Lilly}, {Flores}, {Webb}, {Clements}, {Stevens} \& {Thuan}}{{Waskett}
  et~al.}{2003}]{2003MNRAS.341.1217W}
{Waskett} T.~J.,  {Eales} S.~A.,  {Gear} W.~K.,  {Puchnarewicz} E.~M.,  {Lilly}
  S.,  {Flores} H.,  {Webb} T.,  {Clements} D.,  {Stevens} J.~A.,    {Thuan}
  T.~X.,  2003, mnras, 341, 1217

\bibitem[\protect\citeauthoryear{{Wold}, {Lacy}, {Lilje} \& {Serjeant}}{{Wold}
  et~al.}{2000}]{2000MNRAS.316..267W}
{Wold} M.,  {Lacy} M.,  {Lilje} P.~B.,    {Serjeant} S.,  2000, mnras, 316, 267

\bibitem[\protect\citeauthoryear{{Wold}, {Lacy}, {Lilje} \& {Serjeant}}{{Wold}
  et~al.}{2001}]{2001MNRAS.323..231W}
{Wold} M.,  {Lacy} M.,  {Lilje} P.~B.,    {Serjeant} S.,  2001, mnras, 323, 231

\bibitem[\protect\citeauthoryear{{Xu}, {Livio} \& {Baum}}{{Xu}
  et~al.}{1999}]{1999AJ....118.1169X}
{Xu} C.,  {Livio} M.,    {Baum} S.,  1999, aj, 118, 1169

\bibitem[\protect\citeauthoryear{{Yee} \& {L{\' o}pez-Cruz}}{{Yee} \& {L{\'
  o}pez-Cruz}}{1999}]{1999AJ....117.1985Y}
{Yee} H.~K.~C.,  {L{\' o}pez-Cruz} O.,  1999, aj, 117, 1985

\end{thebibliography}
\bibliographystyle{mn2e}
\bsp

\end{document}